\pdfoutput=1
\documentclass[multicol]{article}
\usepackage[utf8]{inputenc}
\usepackage[T1]{fontenc}
\usepackage[affil-it]{authblk}
\usepackage{natbib}
\usepackage{graphicx}
\usepackage{longtable}
\usepackage{epstopdf}
\usepackage{multicol, caption}
\usepackage[margin=0.8in]{geometry}
\usepackage{sectsty}
\sectionfont{\fontsize{10}{10}\selectfont \centering}
\subsectionfont{\fontsize{10}{10}\selectfont \centering}
\usepackage{array}
\newcolumntype{L}[1]{>{\raggedright\let\newline\\\arraybackslash\hspace{0pt}}m{#1}}
\newcolumntype{C}[1]{>{\centering\let\newline\\\arraybackslash\hspace{0pt}}m{#1}}
\newcolumntype{R}[1]{>{\raggedleft\let\newline\\\arraybackslash\hspace{0pt}}m{#1}}

\title{\textbf{\Large Early deafness leads to re-shaping of global functional connectivity beyond the auditory cortex}}

\newcommand{\beginsupplement}{%
        \setcounter{table}{0}
        \renewcommand{\thetable}{S\arabic{table}}%
        \setcounter{figure}{0}
        \renewcommand{\thefigure}{S\arabic{figure}}%
     }

\author[1,2,+]{Kamil Bonna}
\author[1,+]{Karolina Finc}
\author[3]{Maria Zimmermann}
\author[3]{Łukasz Bola}
\author[3]{Piotr Mostowski}
\author[4]{Maciej Szul}
\author[4]{Paweł Rutkowski}
\author[1,2]{Włodzisław Duch}
\author[5]{Artur Marchewka}
\author[6]{Katarzyna Jednoróg}
\author[3,*]{Marcin Szwed}

\affil[1]{Centre for Modern Interdisciplinary Technologies, Nicolaus Copernicus University in Toruń, Toruń, 87-100, Poland}
\affil[2]{Institute of Physics, Faculty of Physics, Astronomy and Informatics, Nicolaus Copernicus University in Toruń, Toruń, 87-100, Poland}
\affil[3]{Department of Psychology, Jagiellonian University, Kraków, 30-060, Poland}
\affil[4]{Section for Sign Linguistics, Faculty of Polish Studies, University of Warsaw, Warsaw, 00-927, Poland}
\affil[5]{Laboratory of Brain Imaging, Neurobiology Center, Nencki Institute of Experimental Biology, Polish Academy of Sciences, Warsaw, 02-093, Poland}
\affil[6]{Laboratory of Psychophysiology, Neurobiology Center, Nencki Institute of Experimental Biology, Polish Academy of Sciences, Warsaw, 02-093, Poland}

\affil[*]{m.szwed@uj.edu.pl}

\affil[+]{these authors contributed equally to this work}

\begin{document}

\maketitle
 \begin{abstract}
 Early sensory deprivation such as blindness or deafness shapes brain development in multiple ways. While it is established that deprived brain areas start to be engaged in the processing of stimuli from the remaining modalities and in high-level cognitive tasks, some reports have also suggested the possibility of structural and functional changes in non-deprived brain areas. We compared resting-state functional network organization of the brain in early-deaf adults and hearing controls by examining global network segregation and integration. Relative to hearing controls, deaf adults exhibited an altered modular organization, with regions of the salience network coupled with the fronto-parietal network. They showed weaker connections between auditory and somatomotor regions, stronger coupling between the fronto-parietal network and several other large-scale networks (visual, memory, cingulo-opercular and somatomotor), and an enlargement of the default mode network. Their overall functional segregation of brain networks was also lower. Our findings suggest that brain plasticity in deaf adults is not limited to changes in auditory cortex but additionally alters the coupling between other large-scale networks. These widespread functional connectivity changes may provide a mechanism for the superior behavioral performance of the deaf in visual and attentional tasks. 
    \end{abstract}

\begin{center}
\end{center}

\vspace{0.5cm}
\begin{multicols}{2}

\section*{INTRODUCTION}

The lack of input from one sensory modality profoundly impacts brain development \citep{rauschecker1995processing,bavelier2002cross, merabet2010neural}. First, a large body of data shows that the deprived brain regions become involved in the processing of stimuli from remaining modalities. Following deafness, the auditory cortex becomes involved in tactile \citep{lessard1998early, karns2012altered, auer2007vibrotactile} and visual \citep{petitto2000speech, finney2001visual,finney2003visual,DEWEY201555,bola2017task,scott2014enhanced} processing, while following blindness the visual cortex becomes involved in tactile \citep{held1996activation, sadato1998neural} and auditory \citep{alho1993auditory,kujala1997faster,lessard1998early,kujala2005role} processing. Second, sensory deprived areas can also become engaged in higher-level cognitive tasks such as sign language processing \citep{nishimura1999sign,macsweeney2002speechreading,sadato2004age}, speechreading \citep{macsweeney2001dispersed,capek2008cortical}, visual attention \citep{bavelier2001impact} and visuo-spatial working memory \citep{ding2015cross} in the deaf and speech processing \citep{roder2002speech, burton2002adaptive,burton2003visual,amedi2003early,burton2006reading,bedny2015visual}, syntax processing \citep{bedny2015visual,lane2015visual,bedny2017evidence}, mathematical thinking \citep{kanjlia2016absence} and verbal memory \citep{amedi2003early} in the blind. This functional reorganization has been associated with anatomical changes in sensory deprived areas \citep{emmorey2003morphometric,jiang2009thick}. 

Alterations in the brain structure and function of blind or deaf individuals are not restricted to sensory deprived areas. In the blind, besides a volume reduction in the visual cortex, several studies have reported an enlarged auditory cortex \citep{elbert2002expansion}, hippocampus \citep{fortin2008wayfinding} and frontal brain areas \citep{lepore2009pattern}. In the deaf, researchers found an increased volume of the frontal areas \citep{lepore2010brain} and of the insula \citep{allen2008morphology}. Moreover, deaf individuals showed an increased recruitment of multimodal parietal and occipital areas during performance of attention tasks \citep{neville1987attention1, bavelier2000visual,bavelier2001impact,armstrong2002auditory,finney2003visual},  and an increased recruitment of the insula, anterior cingulate and thalamus during verbal memory tasks \citep{bavelier2008ordered}. As the presented evidence suggests, changes caused by deprivation of sensory input are not limited to the corresponding domain-specific cortices, e.g. the primary auditory cortex in deafness. However, the possible impact of these changes on whole-brain network organization has not received adequate attention.

Resting-state fMRI (rsfMRI) offers a solution to investigate whole-brain functional network organization with no explicit task requirements \citep{van2010exploring}. Using rsfMRI data, one can estimate functional connectivity (FC) between different brain areas by measuring the temporal dependence of the low-frequency ($<$ 0.1 Hz) MRI signal fluctuations among them \citep{friston1993functional,biswal1995functional}. Studies on early-blind subjects reported both enhancement and attenuation of functional connections. In the blind, FC enhancement was found between the visual and language networks \citep{liu2007whole,bedny2011language} and between the visual cortex and regions associated with memory and cognitive control \citep{burton2014resting}; weakened connections were found mostly between the visual and somatosensory networks, and between the motor and auditory networks \citep{liu2007whole}. Studies on deaf individuals found an increased FC of the right auditory cortex (superior temporal gyrus, STG) with key nodes of the salience network such as the anterior insula and the dorsal anterior cingulate cortex (dACC) \citep{ding2016enhanced}. Increased resting-state FC was also reported between the right superior parietal gyrus (rSPG) and the right insula, and between the middle temporal gyrus and the posterior cingulate gyrus \citep{li2016functional}; this indicates that changes in connectivity can occur beyond the deprived auditory cortex.

Graph theoretical measures of rsfMRI functional connectivity provide an effective way to describe and understand the organizational features of brain networks in health, disease and throughout development \citep{he2010graph}. To the best of our knowledge, only one study so far has used graph theory measures on rsfMRI data in early-deaf adults \citep{li2016functional}. In the whole brain analysis, Li and colleagues reported hub regions in the frontal and parietal cortices in the deaf but not in the control group. However, the authors did not examine network integration or modular organization, measures commonly applied to characterize plasticity during development \citep{chen2015development} 
and following brain injury \citep{nakamura2009resting}. Brain network integration can be interpreted as the overall efficiency in exchanging information, whereas modular organization describes the propagation and processing of local information. Thus, the question of the level of integration and segregation of functional networks in early-deaf adults remains unanswered. 

The goal of the present study was to examine differences between the whole-brain functional networks of early-deaf and hearing adults. We sought to compare whole-brain connectivity patterns between these two groups, focusing on differences in the strength of functional association between individual brain regions as well as graph measures of the segregation and integration of the entire network \citep{sporns2013network, van2010exploring}. We hypothesized that deaf adults have increased integration of multiple cortical areas. Specifically, we expected to find differences in the connection strength beyond the auditory network. 

\section*{METHOD}

\subsection*{Participants}

Twenty-five early-deaf subjects (15 females; M$_{age}$ = 27.8 $\pm$ 5.2; range 19-37 years) and 29 hearing subjects (16 females; M$_{age}$ = 27.2 $\pm$ 4.7; range 19-37 years) participated in the study. All subjects were right-handed with normal or corrected to normal vision and no neurological or psychiatric diseases. Four deaf subjects and eight hearing subjects were excluded from further analyses due to excessive motion (more than 10\% of outlier scans identified by a scrubbing procedure; see Data Processing section) or image acquisition errors. After exclusion, the deaf group consisted of 21 subjects (14 females; M$_{age}$ = 26.6 $\pm$ 4.8; range 19-37 years) and hearing group of 21 subjects  (14 females; M$_{age}$ = 26.6 $\pm$ 5.2; range 19-37 years). The groups did not differ in age, sex or years of education. The etiology of deafness was either genetic (hereditary deafness) or pregnancy-related (maternal disease or drug side effects). The mean hearing loss was 100.2 dB (range 70–120 dB) for the left ear and 101.4 dB (60–120 dB) for the right ear. All subjects had some experience with hearing aids (currently or in the past) but did not rely on them on a daily basis. All subjects were proficient users of Polish Sign Language (polski język migowy, PJM, a natural visual-gestural language used by the deaf community in Poland; see Table S1 for details). 
Written informed consent was obtained from each participant. During the study, deaf subjects were assisted by a PJM interpreter. The research project was approved by the Committee for Research Ethics of the Institute of Psychology of Jagiellonian University, in accordance with the Declaration of Helsinki. 

\subsection*{Data acquisition}

Neuroimaging data were collected using Siemens MAGNETOM Tim Trio 3T scanner with a 32-channel head coil (Erlangen, Germany). Resting-state functional images covering the whole brain were acquired with a gradient-echo planar imaging (EPI) sequence (33 axial slices in interleaved ascending order; repetition time (TR) = 2190 ms; echo time (TE) = 30 ms, flip angle = 90$^\circ$; field of view (FOV) = 192; matrix size = 64 $\times$ 64; slice thickness = 3.6 mm; voxel size = 3 $\times$ 3 $\times$ 3.6 mm). During the 10-minute resting-state run, 282 volumes were obtained for each subject. Participants were instructed to relax and focus on the fixation point displayed on the screen. Communication with deaf subjects in the scanner was provided in PJM via webcam video.
High-resolution T1-weighted images were acquired using a magnetization-prepared rapid acquisition gradient echo (MPRAGE) sequence (176 slices; TR = 2530 ms; TE = 3.32; flip angle = 7$^\circ$; FOV = 256; voxel size = 1 $\times$ 1 $\times$ 1 mm). 

\subsection*{Data processing}

Neuroimaging data were preprocessed using the SPM12 toolbox (Wellcome Department of Imaging Neuroscience, Institute of Neurology, London, UK) running on MATLAB 8.3 (R2014a) (Mathworks, Natick, MA). First, resting-state functional images were corrected for acquisition time (slice-timing) and spatially realigned to the mean image using rigid body registration. Next, outlier scans with a mean signal higher than 3 SD and frame-displacement (FD) higher than 0.5 mm were identified using the Artifact Detection Toolbox (ART; \url{http://www.nitrc.org/projects/artifact_detect/}). Only subjects with less than 10\% of outlier scans detected were included in the subsequent analysis. There was no significant difference between the deaf and the control group in the mean motion ($t$(39.85) = -0.37; $p$ = 0.71) and the number of outlier scans detected ($t$(31.93) = -0.62; $p$ = 0.54). 

Then, the structural image was coregistered to the first functional volume and functional images, gray matter, white matter (WM), and cerebrospinal fluid were normalized to the MNI  space (voxel size: 2 $\times$ 2 $\times$ 2 mm) using a unified normalization-segmentation algorithm \citep{ashburner2005unified}.

Further data processing for the purpose of functional connectivity analysis was performed using the CONN Functional Connectivity Toolbox v. 17.f \citep{whitfield2012conn} (\url{http://www.nitrc.org/projects/conn/}). The anatomical component correction (aCompCor) strategy was used to estimate and remove physiological noise \citep{behzadi2007component}. The principal components of the subject-specific WM, cerebrospinal fluid, as well as outlier scans detected by the ART procedure and the six rigid-body motion parameters (and their first level temporal derivatives) were removed in covariate regression analysis \citep{whitfield2012conn}. Finally, the resting-state time series were filtered using a 0.008–0.09 Hz band-pass filter to remove the effect of high-frequency noise and low-frequency drift. 

\subsection*{Network construction}

A brain parcellation containing 264 regions of interests (ROIs) provided by functional neuroimaging data meta-analysis  \citep{power2011functional} was selected to construct correlation matrices for the purpose of the whole-brain network analysis. This brain parcellation was extensively validated on other datasets and was used to divide the 264 ROIs into 13 large-scale networks (LSNs) \citep{power2011functional,cole2014intrinsic}. Each ROI was modeled as a 10 mm diameter sphere centered around the coordinates listed by Power et al. \citep{power2011functional}. Six ROIs (four cerebellar ROIs and two ROIs covering the inferior temporal gyrus) were excluded from analysis due to incomplete coverage of the brain in some participants. Denoised functional time series were extracted from the remaining ROIs and Pearson’s correlation coefficients were calculated for each pair of regions. This resulted in one 258 $\times$ 258 correlation matrix for each participant. Finally, Fisher’s transformation was used to normalize Pearson’s correlation coefficients into z-scores.

\subsection*{Edge-wise comparisons}
We aimed to identify functional connections for which the connection strength was either increased or decreased in the deaf group in comparison to the control group. We used a mass univariate approach by independently testing each of the m = 33153 functional connections for difference in connectivity strength between deaf and control subjects with a two-tailed t-test. Then, we estimated associated p-values and corrected with false discovery rate (FDR), using bootstrap method with N$_{per}$ = 10000 permutations \citep{genovese2002thresholding}.

\subsection*{Whole-brain graph measures}

In order to determine if functional brain networks exhibit different topological properties in hearing versus deaf subjects, for each participant we created a weighted, undirected graph by proportionally thresholding the functional connectivity matrix to retain the top 10-25\% functional connections (with a step of 5\%). Here we present the results for the remaining 25\% of connections. As graph metrics depend on network cost (sum of connection strengths) \citep{rubinov2011weight}, we normalized them – on a subject level – against a set of randomly rewired null networks \citep{maslov2002specificity}. Specifically, for each functional network we created 100 null networks with preserved size and degree distribution, and random topology. Then, to estimate null distributions of network metrics we calculated them for respective set of null networks. Finally, we normalized each functional network metric by dividing it by the mean value of the corresponding null distribution. We focused on two widely used graph metrics of network segregation and integration: modularity and global efficiency. All graph measures were calculated using the Brain Connectivity Toolbox \citep{rubinov2010complex}.

The modularity of a network quantifies the extent to which it can be divided into modules. Informally, module is densely interconnected set of nodes sparsely connected with the rest of the network \citep{newman2006modularity}. For a weighted network, modularity is calculated by maximizing the modularity quality function:

$$Q=\frac{1}{v}\sum_{ij}(A_{ij}-\frac{s_is_j}{v})\delta_{m_i,m_j},$$

where $A_{ij}$ is a weighted connection strength between nodes $i$ and $j$, $v$ is the cost of the network $v=\sum_{ij}A_{ij}$, $s_{i}=\sum_{i}A_{ij}$ is the strength of a node, and $\delta_{m_i,m_j}$ is the Kronecker delta that equals 1 when nodes i and j belong to the same community and 0 otherwise. To find the community structure by maximizing $Q$, we ran the Louvain algorithm \citep{blondel2008fast} 100 times per network, and considered the division that yielded the highest modularity value.

Global efficiency $E_{glo}$ enabled us to quantify a network integration by measuring the length of shortest paths between pairs of network nodes. In a weighted network the shortest path can be calculated as the path with a smallest sum of inverse weights, since the stronger connections are intuitively associated with more efficient communication. Formally, weighted global efficiency is given by:

$$E_{glo}=\frac{1}{n(n-1)}\sum_{i}\sum_{j,j\neq i}(d_{ij})^{-1},$$

where $d_{ij}$ is shortest weighted path length between $i$ and $j$.

	To test whether graph metrics differ between deaf and hearing groups we used the non-parametric Wilcoxon rank sum test \citep{wilcoxon1945individual}.

\subsection*{Large-scale brain networks}

To assess more general differences of the large-scale brain systems between hearing and deaf participants we examined the modular structure of the functional networks in both groups. For each group, we created a single representative network by averaging connection strengths across subjects. To eliminate insignificant connections in each group-averaged connection matrix we calculated the significance of the connection strength against zero. Assessed p-values were corrected using the false discovery rate (FDR) method for both groups separately \citep{genovese2002thresholding}. Connections that survived thresholding, i.e. those with p$_{FDR} <$ 0.05, were retained in the group-averaged connection matrix. Next, to establish a representative modular structure, we ran the Louvain algorithm 1,000 times for both group-averaged networks and considered runs that produced divisions with the highest modularity value. Finally, we compared the modular structure in the deaf and hearing group with the large-scale network division revealed by resting-state meta-analysis \citep{power2011functional,cole2014intrinsic}. 

In order to quantify our findings, we calculated the overlap coefficient between empirically found modules and well-known large-scale brain systems. Overlap coefficient is a measure of similarity between two overlapping sets. Here, as sets we consider subsets of nodes grouped together in large-scale module. Formally, for two sets A and B overlap coefficient is given as

$$overlap(A,B)=\frac{\left | A\cap B \right |}{min(\left | X \right |,\left | Y \right |)},$$

where $\left| \cdot \right|$ denotes the number of elements of the set. Note that the overlap coefficient equals one for every pair of sets that $A\subseteq B$ or $B\subseteq A$.

\section*{RESULTS}

\subsection*{Edge-wise whole-brain differences between the deaf and hearing}

\begin{figure*} 
    \centering    
    \includegraphics[width=0.8\textwidth]{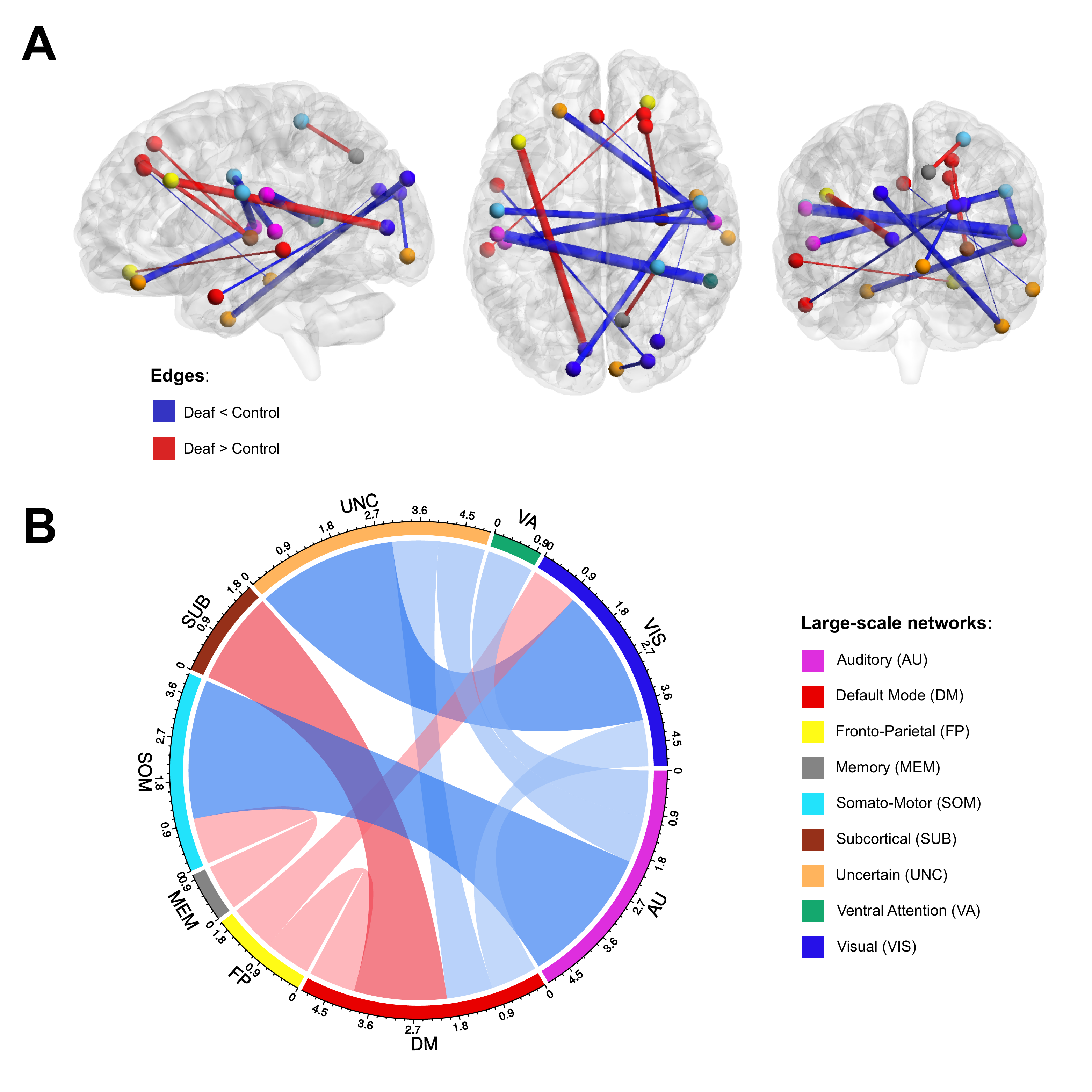}
    \captionof{figure}{Edge-wise functional network differences visualized (A) in brain space and (B) as a chord diagram. (A) Connections that are significantly stronger (red) or weaker (blue) in deaf adults. Edge thickness reflects t-test statistic strength. (B) Chord diagram representing the number of significant edges between different large-scale networks. Red bands represent edges with stronger functional connectivity in the deaf compared to hearing control, while blue bands represent edges with weaker functional connectivity.}
    \label{fig:fig1}
\end{figure*}

We compared the strength of all pairwise functional connections (edges) between 258 ROIs in the deaf versus the control group. These comparisons revealed 10 weaker and 5 stronger connections in early-deaf adults (Fig. \ref{fig:fig1}, FDR corrected $p <$ 0.05). Weaker connections in the deaf relative to the controls were found mostly between the auditory and somatomotor networks, as well as between the visual network and regions not assigned to any large-scale networks. Interestingly, stronger connections in the deaf were found between regions beyond the auditory network. These included two enhanced connections between the default mode network and the subcortical network. Enhanced connections were also found between the fronto-parietal and default-mode networks, between the fronto-parietal and visual networks and between the memory and somatomotor networks (see Fig. \ref{fig:fig1}b for edge counts after large-scale network assignment).

\subsection*{Differences in whole-brain graph measures}
 
Functional brain network topology is believed to support an optimal balance between functional segregation and integration enabling complex network dynamics \citep{tononi1994measure}. These two network features can be captured using two graph theory measures: modularity index for segregation \citep{newman2006modularity} and global efficiency for integration \citep{latora2001efficient}. Here, we tested whether these measures differ between deaf and hearing subjects (Fig. \ref{fig:fig2}). Analysis performed on brain graphs parcellated with 258 functional ROIs revealed significant differences in network modularity ($z-val$ = -2.36; $p$ = 0.019, Wilcoxon rank sum test, see Methods) between the two groups. Whole-brain modularity was lower in deaf participants ($Q_{deaf}$=3.50; $std$($Q_{deaf}$) = 0.31) than in hearing participants ($Q_{control}$ = 3.65; $std$($Q_{control}$) = 0.14 Wilcoxon rank sum test). A significant difference in modularity was also observed for functional networks constructed for all threshold values ($p <$ 0.05). The difference in functional network integration measured as global efficiency ($z-val$ = 1.26; $p$ = 0.21, Wilcoxon rank sum test) was not significant. Taken together, these results imply that functional brain networks in early-deaf adults are less segregated than those in hearing adults. 

\begin{figure*} 
    \centering    
    \includegraphics[width=0.6\textwidth]{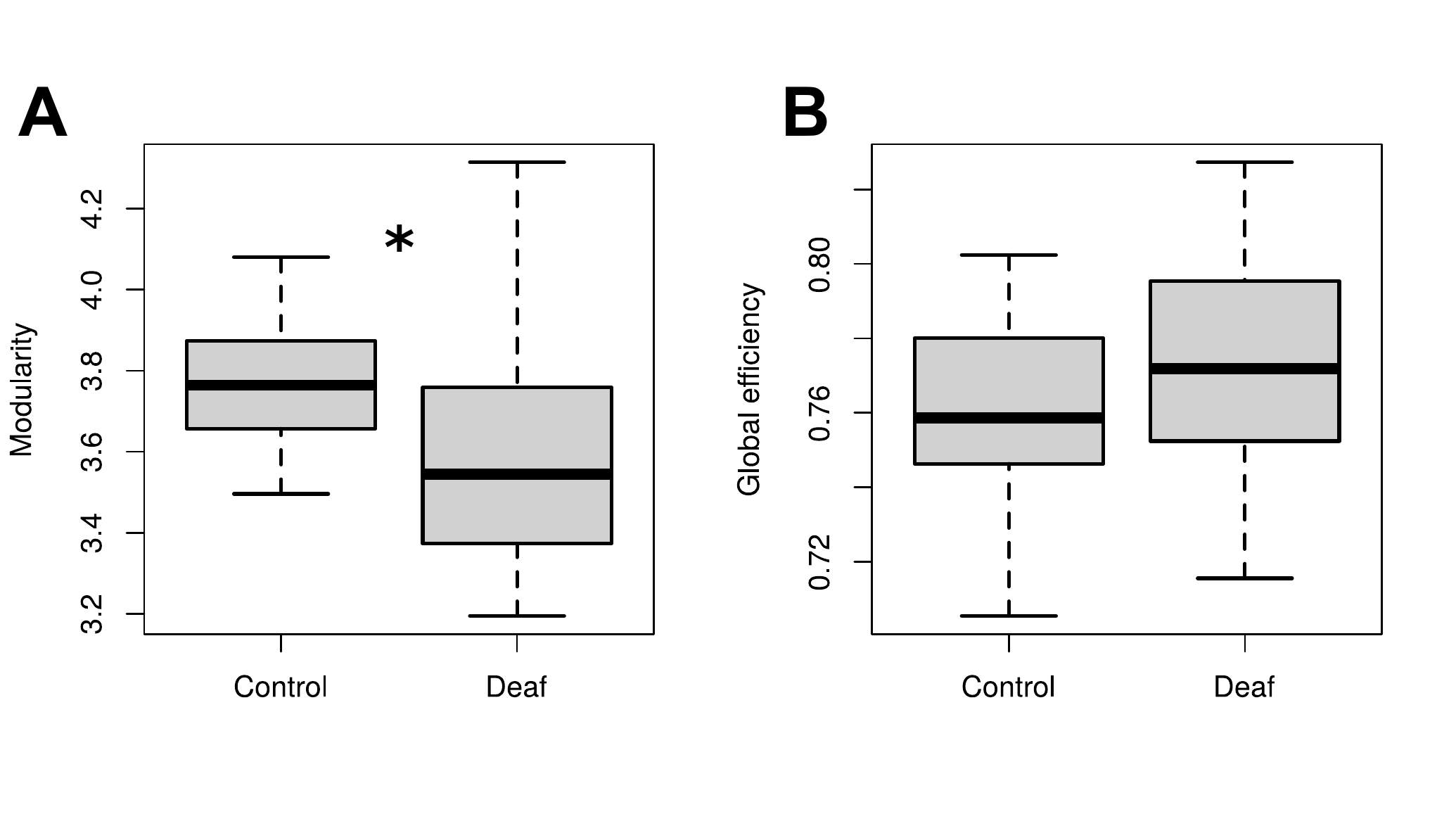}
    \captionof{figure}{Differences in graph measures of cortical segregation and integration between deaf adults and the control group. (A) Difference in network segregation measured as modularity. (B) Difference in network integration measured as global efficiency. Boxplots represent topological values calculated for 25\% threshold. *$p <$ 0.05}
    \label{fig:fig2}
\end{figure*}

\subsection*{Group-average modular organization}

\begin{figure*} 
    \centering    
    \includegraphics[width=0.6\textwidth]{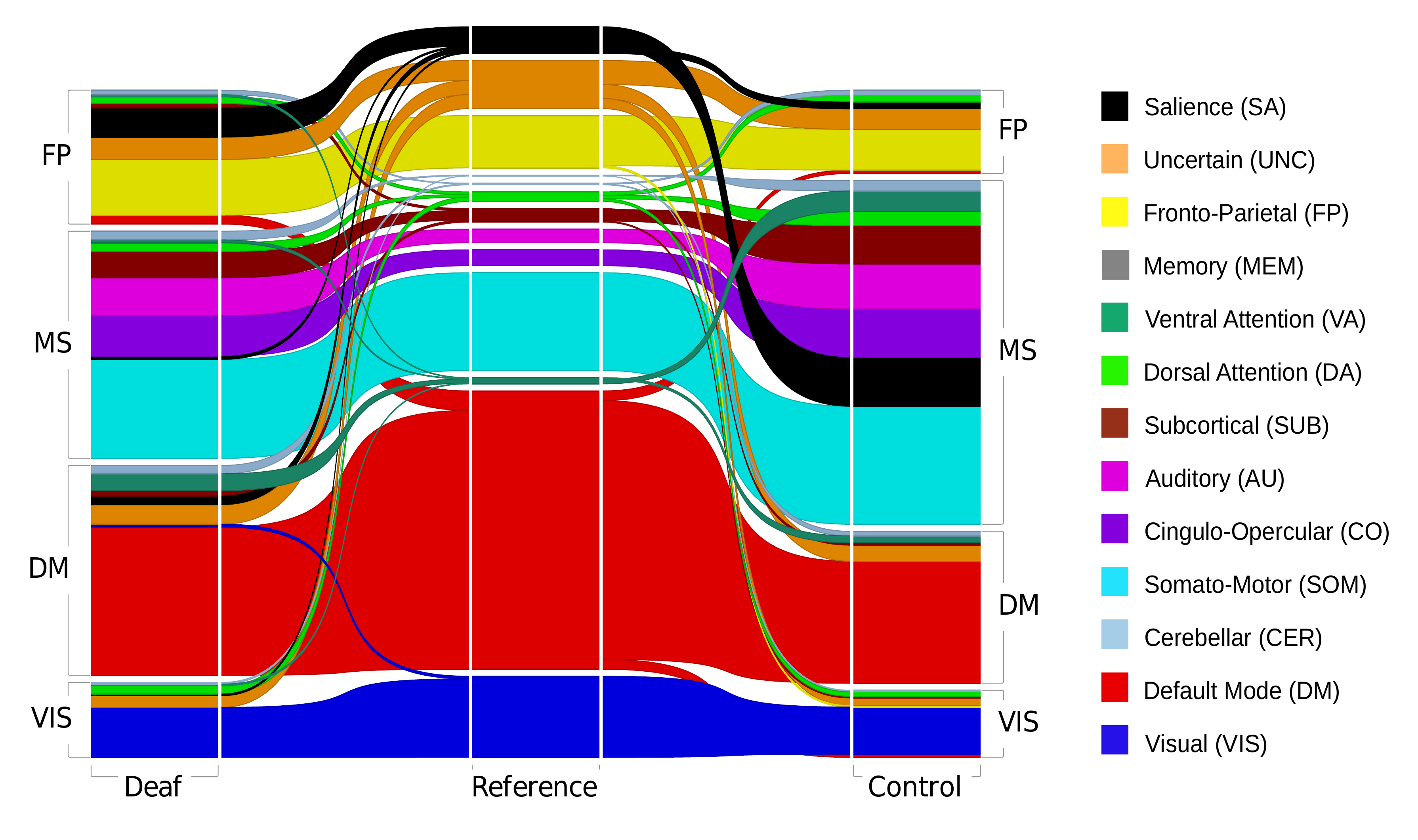}
    \captionof{figure}{An alluvial diagram representing the segregation of group-averaged networks using a data-driven approach in the deaf (left side of the diagram) and the control group (right side of the diagram). This segregation is then compared against an a priori segregation into 13 well-known networks based on meta-analysis studies \cite{power2011functional}, shown in the middle column and described in the right-hand side legend. Note that salience nodes (black) are part of the fronto-parietal (FP) module in the deaf group but fall into the multi-system (MS) module in the control group. In addition, the ventral-attention nodes (dark green) are part of the MS module in the control group, but in the deaf group they are part of the default mode module (DM). The composition of the last, visual module (VIS) is highly consistent in both groups.}
    \label{fig:fig3}
\end{figure*}

In the analysis that followed we assessed the modular division of the group-averaged networks using a data-driven approach\citep{blondel2008fast} (see Methods). We found that for both groups this approach returned a connectivity structure arranged into four large-scale functional modules (Fig. \ref{fig:fig3}, Fig. S1): the fronto-parietal (FP) module, the multi-system (MS) the default mode module (DM) and the visual module (VIS). 
In both groups we then analyzed the overlap of these four modules with 13 well-known large-scale networks (LSNs) that were defined a priori based on meta-analyses \citep{power2011functional,cole2014intrinsic}) (Fig. \ref{fig:fig3}) by calculating an overlap coefficient between the data-driven modules and all 13 LSNs. In this analysis, an overlap coefficient of 100\% means that a given network (for example, the somatomotor network) is completely included in a given module (for example, the multi-system module).
The first module, the fronto-parietal module, consisted mostly of regions from the fronto-parietal network (overlap in the deaf group ($X^{fp}_{deaf}$, $X^{fp}$) = 100\%; overlap in the control group ($X^{fp}_{control}$, $X^{fp}$) = 96\%). This module had a significantly different composition in the deaf as compared to the control group. In the deaf, the salience network was associated almost exclusively with the fronto-parietal network within the fronto-parietal module (overlap in the deaf group ($X^{fp}_{deaf}$, $X^{sal}$) = 72.2\%). In the hearing group, however, the fronto-parietal module made only a small contribution to the salience network nodes (overlap in the control group ($X^{fp}_{control}$, $X^{sal}$) = 22.2\%), which turned out to be predominantly associated with the multisystem module (Fig. 3, black). 
The second module (referred to here as the multi-system module) was the largest and most diverse module ($|X^{ms}_{deaf}|$ = 78; $|X^{ms}_{deaf}|$ = 99). In the control group, it was composed of the somatomotor, salience, auditory, cingulo-opercular, ventral-attention, subcortical and cerebellum nodes (overlap > 66\%). In the deaf group, however, this module did not include the salience and ventral-attention networks (overlap($X^{ms}_{deaf}$, $X^{sal}$) = 5.5\%; overlap ($X^{ms}_{deaf}$, $X^{va}$) = 11.1\%) which were associated with other modules, i.e. the salience network with the fronto-parietal module, and the ventral attentional network with the default mode module. 
The third module, the default mode module, had a very high overlap with the default mode network (overlap in the deaf group ($X^{dm}_{deaf}$, $X^{dmn}$) = 93\%, overlap in the control group ($X^{dm}_{control}$, $X^{dm}$) = 93\%). It consisted of 75 nodes in the deaf group and 66 nodes in the control group. The default mode module was larger in the deaf group as a result of large contribution from the ventral-attention (overlap the deaf group ($X^{dm}_{deaf}$,$X_{va}$) = 66.7\%) and memory network nodes (overlap in the deaf group ($X^{dm}_{deaf}$,$X^{mem}$) = 60\%), which, as mentioned previously, in the hearing group were mostly part of the multi-system module.
The last module, the visual module, was the most consistent in both groups (overlap($X^{vis}_{deaf}$, $X^{vis}_{control}$) = 88.6\%; $|X^{vis}_{deaf}|$ = 44; $|X^{vis}_{control}|$ = 45). In both groups it was composed primarily from visual network nodes. In agreement with the previous modularity analysis, we also found that the group-averaged functional network was less modular in the deaf group than in the hearing group ($Q_{deaf-av}$ = 0.4571; $Q_{control-av}$ = 0.4748).

\section*{DISCUSSION}

In this study we investigated the whole-brain functional organization differences between early-deaf and hearing adults. Deaf adults exhibited weaker connection strengths, especially between the auditory and the somatomotor networks. Besides changes in the functional connectivity of auditory regions, we also found pronounced connectivity differences between regions located outside of the auditory system. These differences included a stronger functional connectivity between the fronto-parietal network and other large-scale networks (salience, visual, memory, cingulo-opercular and somatomotor, default mode) and between the default mode and the subcortical network in deaf adults. Using graph theoretical measures, we showed that deaf adults had a less segregated (modular) network. We also found different modular organization of functional networks in deaf subjects. Differences were pronounced for the salience and ventral-attention systems: in the control group they were part of a multi-system module, but in the deaf they were coupled with the fronto-parietal and default-mode modules. These results suggest that compensatory brain plasticity in sensory loss is a combination of changes in the sensory-deprived brain areas themselves and changes beyond this in non-deprived brain areas.

\subsection*{Reduced functional connectivity between auditory and somatomotor areas}

In the deaf, we observed reduced functional connectivity between auditory and somatomotor areas (Fig. \ref{fig:fig1}). Previous results showed cross-modal plasticity of the auditory cortex and its engagement in tactile \citep{auer2007vibrotactile,levanen1998vibration, karns2012altered} and visual processing \citep{petitto2000speech,finney2001visual,finney2003visual,scott2014enhanced, DEWEY201555,bola2017task}. One may intuitively expect that this cross-modal plasticity could be expressed in enhanced functional connections between the auditory and somatomotor networks. However, functional connectivity results are not always consistent with results of task-based fMRI studies, and these two approaches might reveal different properties of brain organization \citep{gratton2016evidence, khambhati2017modeling}. While task-related activations of the auditory cortex in deaf individuals or the visual cortex in blind individuals during tactile processing were consistently found in previous studies \citep{levanen1998vibration,karns2012altered,held1996activation, sadato1998neural, auer2007vibrotactile}, none of the existing functional connectivity studies have reported increased connectivity between the somatomotor cortex and sensorially deprived areas in deaf or blind individuals \citep{liu2007whole,burton2014resting,ding2016enhanced,li2016functional}. Moreover, some studies have demonstrated that blind individuals have weakened functional connections between the visual and somatosensory and motor networks, which is strongly consistent with our results on deaf individuals \citep{liu2007whole}. Recent evidence revealed a very similar effect of weakened connectivity between somatomotor and deprived auditory cortices in deaf cats \citep{butler2018cortical}. Liu and colleagues \citep{liu2007whole} interpreted their finding of weakened connectivity between sensory deprived areas and the somatomotor network in terms of general loss hypothesis \citep{pascual2005plastic}. According to this hypothesis, the functional organization of the sensory deprived brain may be generally disrupted because of the lack of sensory input. However, many studies on early sensory deprivation do not support this notion \citep{bavelier2002cross,merabet2005blindness} \citep{theoret2004behavioral,pascual2005plastic}. Here, we propose a different explanation for our findings in terms of the neural efficiency hypothesis \citep{neubauer2009intelligence}. According to this hypothesis, less engagement of certain brain areas during task processing may indicate that the task is performed more automatically, with less energy consumption. In the context of functional connectivity, learning a new skill can be associated with a reduced connectivity between areas associated with the trained domain \citep{kelly2014strengthening,bassett2015learning, yoo2013tool,wang2016exploring}. For example, Yoo et al.\citep{yoo2013tool} reported decreased functional connectivity between the parietal cortex and the motor system after 8 weeks of practicing using chopsticks with a non-dominant hand. Bassett et al. \citep{bassett2015learning} found that training of a visuomotor task is associated with reduced connectivity between the visual and motor networks, suggesting that these systems are autonomous in relation to task automatization. Also, Wang et al. \citep{wang2016exploring} reported reduced functional connectivity of the somatomotor system in world-class gymnasts. In the case of deaf individuals, the trained skill may be related to the use of sign language to communicate and to superior tactile processing. Taken together, the reduced functional connectivity between the auditory and somatomotor areas in early-deaf adults may be linked to more efficient, automated sensorimotor processing rather than a general loss of connectivity.

\subsection*{Increased fronto-parietal connectivity in deafness}

Besides the connectivity decreases outlined above, deaf subjects exhibited strengthened interconnections, notably with the frontoparietal system. Edge-wise analysis  revealed an increased coupling between the fronto-parietal network and visual areas (Fig. \ref{fig:fig1}). This coupling might support the higher need for visual attention resources in the deaf. While deaf subjects consistently outperform hearing subjects in several visual tasks \citep{scott2014enhanced,DEWEY201555} this occurs almost exclusively under conditions of high attentional load \citep{heimler2017multisensory}. We speculate that enhanced connectivity between the sensory and fronto-parietal networks may provide the neural basis for visual compensation mechanisms. The altered functional role of the fronto-parietal system and the enhanced visual–fronto-parietal interconnections may constitute the neural basis for the congenitally deaf’s superior performance in both sensory attention and visual working memory.

\subsection*{Decreased modularity of functional networks in deafness}

Modularity of whole-brain functional networks was lower in deaf subjects compared to hearing controls, thus indicating disrupted boundaries between functionally specialized systems following early deafness (Fig. \ref{fig:fig2}). Our findings provide the first evidence of an altered modular organization of functional networks in sensory deprived subjects. Several studies reported disrupted modular organization associated with healthy aging \citep{song2012age, geerligs2014brain}, adolescence \citep{fair2009functional}, childhood-onset schizophrenia \citep{alexander2010disrupted} and autism spectrum disorder \citep{rudie2013altered}. Other studies reported increased modularity in patients with major depressive disorder \citep{ye2015changes} or attention-deficit/hyperactivity disorder \citep{lin2014global}. Here, for the first time we show that sensory deprivation may reduce modularity at the global brain network level. These results suggest that sensory deprivation can lead to a blurring of the lines between specialized brain subsystems, while network integration (measured as global efficiency) remaining at the same level as in normally developing individuals. 

\subsection*{Coupling of salience and fronto-parietal networks in deafness}

Our study found that that the salience network was associated with the fronto-parietal network in the deaf group, but not in the control group. Consistently with our results, in congenitally and early blind subjects the salience network has previously been shown to exhibit stronger resting-state functional connectivity with fronto-parietal regions than in the sighted population \citep{wang2014altered}. The effect revealed in our study may thus be a general consequence of early sensory deprivation.

The salience network is responsible for identifying behaviorally important stimuli, forwarding them to the executive functions network, and mediating higher order cognitive processes \citep{seeley2007dissociable}. Its functional association with fronto-parietal structures plays a role in working memory processing. Activity of the salience network is gradually enhanced with increased working memory load, and this enhancement correlates positively with working memory task performance \citep{liang2015topologically}. It can be therefore inferred that its strengthened association with the fronto-parietal module in deafness reflects the enhanced working memory abilities reported in early sensory deprivation.

In line with this interpretation, task-based studies reveal the specifically important role the salience network can play in deafness for working-memory-related functions. When compared with the hearing, deaf subjects recruit the salience network more strongly for short-term verbal memory tasks \citep{bavelier2008ordered} and exhibit stronger functional connectivity between salience and auditory structures when processing a visual working memory task \citep{ding2016enhanced}. Additionally, deaf subjects have increased gray and white matter within the salience network \citep{allen2008morphology}, and this structural reinforcement has been suggested to contribute to sign language processing \citep{kassubek2004involvement}. 

\subsection*{Altered connectivity of default mode network in deafness}

Our results show differences in the default mode network (DMN) connectivity in the deaf. Whole-brain network analysis revealed that in the deaf the DMN has strengthened connections with the subcortical network and the fronto-parietal network, and a weakened connection with the visual system (Fig. \ref{fig:fig1}). Moreover, we found that the default mode module in the deaf includes the ventral attention system. In the hearing, in contrast, the ventral attention system is coupled with the multi-system module (Fig. \ref{fig:fig3}).

The exact role of the DMN is debatable. Some studies provide evidence that it is associated with internally directed cognitive processing such as mind-wandering or autobiographical memory \citep{buckner2008brain}. The DMN is also often referred to as a task-negative network due to its anticorrelation with networks related to attentional processing \citep{fox2005human}. However, a new wave of research provides evidence for an integrative role of the DMN which may be crucial for higher cognitive functions \citep{vatansever2015default, margulies2016effects, finc2017transition}). In line with this research, stronger connectivity between the DMN and the subcortical and fronto-parietal networks may suggest that the DMN is engaged in network integration that is necessary to compensate for sensory deficits.
    
The ventral attention system is typically recruited by infrequent or unexpected events that are behaviorally relevant, and has been implicated in stimulus-driven, involuntary attentional control \citep{corbetta2002control}. Therefore, it is plausible that its closer association with the DMN in the deaf corresponds to an easier and faster transition between resting state and the action in response to the unexpected input. The ability to shift quickly from rest to action seems to be particularly adaptive in an environment lacking auditory input, as one must deal with a constrained field of view and higher latency in response time for vision. 

Compensatory mechanisms lead deaf people to outperform hearing individuals in certain visual tasks, especially when the location or the exact time of the onset of the stimulus is unknown \citep{bavelier2006deaf}, or when the stimulus appears outside the central visual field. The lack of auditory signal is compensated for in the deaf by enhanced peripheral visual attention \citep{lore1991central,neville1987attention1,stevens2006neuroplasticity}. These effects make deaf subjects more distractible by peripheral visual input \citep{proksch2002changes}, which may enable them to detect unexpected stimulus more quickly and respond to unpredicted cues in sign language. On the more general level, deaf subjects manifest consistently lower response time to visual input across a variety of visual tasks \citep{pavani2012visual}. The enhanced coupling between resting-state DMN and the ventral attention system in the deaf could thus reflect their general higher reactivity to visual stimuli in deafness as well as more specific capacities in visual attention. 

Overall, our results show substantial differences in the functional brain network organization between early-deaf and hearing adults. We have shown that deaf adults have reduced coupling between the auditory and the visual cortex. However, we also found multiple differences in functional connectivity beyond the auditory network, including the fronto-parietal, default-mode and salience networks. These results suggest that changes in brain connections related to sensory deprivation are not limited to the deprived cortices, but manifest themselves in altered connectivity across the entire brain network.

\section*{Acknowledgements}

Study supported by National Science Centre grants no 2015/19/B/HS6/01256 and 2016/21/B/HS6/03703 to M. S.,  P. M. and P. R. were supported under the National Programme for the Development of Humanities of the Polish Ministry of Science and Higher Education (0111/NPRH3/H12/82/2014). K. F. was supported by National Science Centre (2017/24/T/HS6/00105) and and Foundation for Polish Science, Poland (START 23.2018). The study was conducted with the aid of CePT research infrastructure purchased with funds from the European Regional Development Fund as part of the Innovative Economy Operational Programme, 2007–2013. We thank Karolina Dukała for administrative assistance, and Michael Timberlake for language editing.

\section*{Author contributions statement}
K.B. and K.F. analyzed data;  K.B., K.F, M.Z., and M.S. wrote manuscript; Ł.B., M.Z., and M.S. designed research
and Ł.B., M.Z., P.M., K.J., A.M., P.R., and M.S. performed research; W.D. reviewed manuscript.

\section*{Additional information}
\subsection*{Data availability}
The datasets generated and analysed during the current study are available from the corresponding author on request.
\subsection*{Competing interests}
The authors declare no competing interests.

\bibliographystyle{apalike}
\bibliography{references}

\end{multicols}

\newpage
\beginsupplement
\section*{SUPPLEMENTARY INFORMATION}
\label{Suplementary Information}

\begin{center}
\textbf{Table S1.} Characteristics of deaf participants.    
\end{center}
\begin{longtable}{c|C{1cm}|C{0.8cm}|L{2cm}|L{2cm}|L{1.5cm}|C{2cm}|C{1.5cm}|C{1.5cm}}
SubID & Sex & Age & Cause of deafness & Hearing loss (left ear/right ear/mean) & Hearing aid use & How well the subject understands speech with hearing aid & Native language (oral/sign) & Languages primarily used at the moment of the experiment \\
\hline	
\rule{0pt}{5ex}
Sub01 & F   & 30  & Hereditary deafness                 & 110/90/100 dB                          & Uses currently   & Moderately                                               & Sign                        & Sign                                                     \\
\hline	
\rule{0pt}{5ex}
Sub02 & M   & 27  & Maternal disease/ drug side effect  & 120/90/105 dB                          & Used in the past & Moderately                                               & Sign                        & Sign                                                     \\
\hline	
\rule{0pt}{5ex}
Sub04 & M   & 23  & Hereditary deafness                 & avg: 120 dB, profound               & Uses currently   & Moderately                                               & Sign                        & Sign \& oral                                             \\
\hline	
\rule{0pt}{5ex}
Sub05 & M   & 27  & Hereditary deafness                 & avg: 90-119 dB, severe              & Used in the past & Poorly                                                   & Sign                        & Sign                                                     \\
\hline	
\rule{0pt}{5ex}
Sub06 & M   & 27  & Hereditary deafness                 & avg: 90-119 dB, severe               & Used in the past & Poorly                                                   & Oral                        & Sign \& oral                                             \\
\hline	
\rule{0pt}{5ex}
Sub07 & F   & 27  & Hereditary deafness                 & avg: 120 dB or more, profound       & Used in the past & Poorly                                                   & Oral                        & Sign \& oral                                             \\
\hline	
\rule{0pt}{5ex}
Sub08 & F   & 27  & Hereditary deafness                 & avg: 90-119 dB, severe               & Used in the past & Poorly                                                   & Sign                        & Sign                                                     \\
\hline	
\rule{0pt}{5ex}
Sub09 & M   & 27  & Hereditary deafness                 & 120/120/120 dB                         & Used in the past & Poorly                                                   & Sign                        & Sign                                                     \\
\hline	
\rule{0pt}{5ex}
Sub10 & F   & 32  & Hereditary deafness                 & 89/80/85 dB                            & Uses currently   & Moderately                                               & Sign                        & Sign \& oral                                             \\
\hline	
\rule{0pt}{5ex}
Sub14 & F   & 32  & Maternal disease/ drug side effect  & 105/115/110 dB                         & Uses currently   & Moderately                                               & Oral                        & Sign \& oral                                             \\
\hline	
\rule{0pt}{5ex}
Sub17 & F   & 19  & Hereditary deafness                 & 95/100/98 dB                           & Uses currently   & Moderately                                               & Sign                        & Sign \& oral                                             \\
\hline	
\rule{0pt}{5ex}
Sub18 & M   & 27  & Hereditary deafness                 & 94/107/101 dB                          & Used in the past & Poorly                                                   & Sign                        & Sign                                                     \\
\hline	
\rule{0pt}{5ex}
Sub19 & F   & 30  & Hereditary deafness                 & 90/90/90 dB                            & Used in the past & Poorly                                                   & Sign                        & Sign \& oral                                             \\
\hline	
\rule{0pt}{5ex}
Sub20 & F   & 25  & Hereditary deafness                 & 70/60/65 dB                            & Uses currently   & Well                                                     & Sign                        & Sign                                                     \\
\hline	
\rule{0pt}{5ex}
Sub21 & F   & 37  & Maternal disease/ drug side effect  & 110/110/110 dB                         & Used in the past & Poorly                                                   & Oral                        & Sign                                                     \\
\hline	
\rule{0pt}{5ex}
Sub22 & F   & 20  & Hereditary deafness                 & 113/115/114 dB                         & Used in the past & Poorly                                                   & Oral                        & Sign \& oral                                             \\
\hline	
\rule{0pt}{5ex}
Sub23 & M   & 19  & Hereditary deafness                 & 90/110/100 dB                          & Uses currently   & Well                                                     & Sign                        & Sign \& oral                                             \\
\hline	
\rule{0pt}{5ex}
Sub24 & F   & 19  & Hereditary deafness                 & 94/103/99 dB                           & Uses currently   & Very well                                                & Sign                        & Sign \& oral                                             \\
\hline	
\rule{0pt}{5ex}
Sub28 & F   & 30  & Hereditary deafness                 & 78/92/85 dB                            & Uses currently   & Poorly                                                   & Sign                        & Sign \& oral                                             \\
\hline	
\rule{0pt}{5ex}
Sub29 & F   & 23  & Maternal disease/ drugs side effect & 102/120/111 dB                         & Uses currently   & Moderately                                               & Oral                        & Sign \& oral                                             \\
\hline	
\rule{0pt}{5ex}
Sub31 & F   & 30  & Maternal disease/ drug side effect  & 100/120/110 dB                         & Uses currently   & Well                                                     & Oral                        & Sign \& oral                                            
\end{longtable}

\begin{figure*}[!htbp] 
    \centering    
    \includegraphics[width=0.85\textwidth]{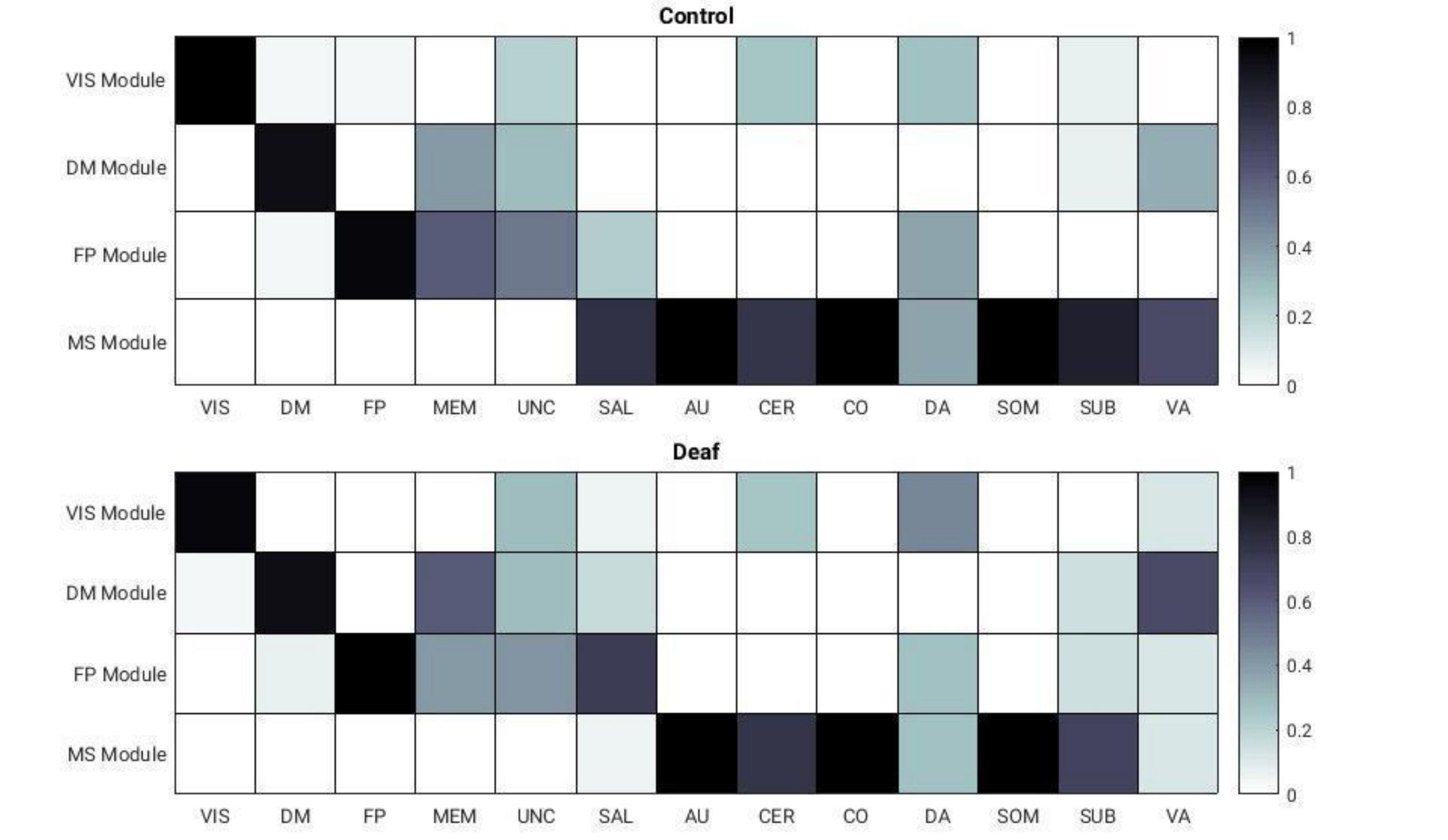}
    \captionof{figure}{Overlap between data-driven representative network modules in hearing and deaf participants and 13 well-known large scale networks (LSNs) (Power et al., 2011). Vertical axis corresponds to four large-scale functional modules discovered by community detection algorithm (see Results). Horizontal axis corresponds to LSNs from Power et al. (2011). Darker colors reflect higher overlap coefficient between pair of modules.}
    \label{fig:fig2}
\end{figure*}

\vspace{1cm}

\end{document}